\documentclass[aps,pre,floatfix]{revtex4}
\usepackage{epsf}
\usepackage{subfigure}
\usepackage{amsmath}
\usepackage{amssymb}
\usepackage{graphicx}
\usepackage{epstopdf}

\newcommand{\mean}[1]{\left \langle #1 \right \rangle}
\newcommand{\parent}[1]{\left( #1 \right)}
\newcommand{\cum}[1]{\left \langle \left \langle #1 \right \rangle \right \rangle}

\begin{document}

(Physical Review E 82, 031124, 2010)

\vspace{1cm}

\title{Thermodynamic large fluctuations from uniformized dynamics}

\author{David Andrieux}

\affiliation{                 
Department of Neurobiology and Kavli Institute for Neuroscience, Yale University School of Medicine, New Haven, CT 06510, USA\\
and\\
Center for Nonlinear Phenomena and Complex Systems, Universit\'e Libre de Bruxelles, B-1050 Brussels, Belgium.
}


\begin{abstract}
Large fluctuations have received considerable attention as they encode information on the fine-scale dynamics. Large deviation relations known as fluctuation theorems also capture crucial nonequilibrium thermodynamical properties. 
Here we report that, using the technique of uniformization, the thermodynamic large deviation functions of continuous-time Markov processes can be obtained from Markov chains evolving in discrete time. 
This formulation offers new theoretical and numerical approaches to explore large deviation properties. 
In particular, the time evolution of autonomous and non-autonomous processes can be expressed in terms of a single Poisson rate.
In this way the uniformization procedure leads to a simple and efficient way to simulate stochastic trajectories that reproduce the exact fluxes statistics. 
We illustrate the formalism for the current fluctuations in a stochastic pump model.
\end{abstract}

\maketitle

\section{Introduction}

Many natural phenomena are successfully described at the mesoscopic level in terms of Markovian random processes \cite{C43,S76,NP77}. 
Examples of such jump processes range from birth-and-death processes in stochastic chemical kinetics and population dynamics \cite{S76, NP77} to kinetic processes in quantum field theory \cite{W05} and in quantum optics \cite{Lo73}. In some simple systems, such processes can be rigorously derived from the underlying deterministic or quantum dynamics by introducing an appropriate partition of the phase space \cite{G98} or in some scaling limit \cite{S80}. 

The study of these continuous-time Markov processes remains, however, challenging. 
The so-called uniformization technique \cite{J53} has been introduced to help the analysis of such continuous-time processes.
The uniformization procedure transforms a continuous-time Markov process into a discrete-time Markov chain, facilitating all subsequent analysis. The term uniformization comes from the fact that the original continuous-time process can be reinterpreted as involving an homogeneous Poisson process along with transitions described by the derived Markov chain. This scheme is especially used for the study of the transient properties of the dynamics,  
but it can also be applied to study other important quantities such as first-passage times \cite{GM84,S86}. Uniformization has also been used to simulate the behavior of complex systems such as chemical reaction networks \cite{W08, ZWC10} or evolutionary models \cite{RPL07}. 
Interestingly, a similar approached was developed in the context of quantum dynamics \cite{G68,AJS83, BPAJ99, OB05}, where Poisson processes provide a generalization of the Feynman-Kac formula \cite{F48,K49} to quantum systems with discrete internal degrees of freedom. 

Fundamental properties of stochastic and deterministic dynamical systems can be expressed in terms of large deviation functions, which characterize the occurrence of rare fluctuations or extreme events in random systems. 
In this framework trajectories are categorized by dynamical order parameters such as the number of configuration changes. Analogous to the partition function in equilibrium statistical mechanics, the large deviation function is a measure of the number of trajectories accessible to the system. Critical phenomena such as scale invariance of trajectories or dynamical phase transitions can be uncovered from the knowledge of the large deviation function. In addition, it gives access to the statistical properties (averages and fluctuations) of the dynamical order parameters. In this sense large deviations can be said to capture the fine details of the dynamics.

Landford \cite{L73} was the first to formulate equilibrium statistical mechanics in terms of large deviations, where they provide a generalization of Einstein's fluctuation theory and allow the calculation of entropies and free energies.
Large deviations were next considered in nonequilibrium statistical physics by Ruelle and Bowen in their analysis of the dynamical properties of chaotic systems \cite{R78}. This formalism was further developed to relate the dynamical properties of these systems to their transport properties \cite{G98}. 
In this way large deviations provide a rigorous formulation of statistical mechanics, as reviewed in Refs. \cite{O89, T09}. 
More recently, relations known as fluctuation theorems \cite{ECM93, GC95, J97, C99} (see \cite{SPWS08} for a review) revealed that thermodynamical quantities obey symmetry relationships when accounting for the large fluctuations in the time evolution (rare trajectories). 
Large deviation relations thus play a fundamental and unifying role in characterizing the dynamical and statistical properties of equilibrium and nonequilibrium systems {\cite{T09}.

The dissipation rate and the thermodynamic currents play an important role in nonequilibrium statistical thermodynamics \cite{S76, NP77}. The dissipation is related to the irreversible entropy production and the efficiency of free energy conversion into useful work. The thermodynamic currents describe the fluxes of matter or energy flowing through the system. Their response and fluctuation properties are therefore of fundamental interest, especially, for the exploration of nanoscale systems.

In this paper we introduce the uniformization procedure of continuous-time Markov processes and show that it can be applied to recover the large deviation functions. We illustrate the construction for two thermodynamic quantities of interest, the dissipation rate \cite{ECM93, GC95, K98, M99, LS99} and the thermodynamic currents \cite{AG, AG07}. We then analyze the time evolution of autonomous and non-autonomous systems. We obtain a formulation in terms of a single Poisson process, which offers new insights at the theoretical and numerical levels. In particular, we show that the simulation of time-dependent systems such as stochastic pumps can be achieved in a effortless and efficient way.\\

\section{Markov processes and uniformization}

Continuous-time Markov processes are ruled by an evolution 
equation, called the master equation, for the probability to find the system in a coarse-grained state $j$ at time $t$: 
\begin{eqnarray}
\frac{dp_j (t)}{dt} = \sum_i [p_i(t) W_{ij} -  p_j(t) W_{ji} ] \, .
\end{eqnarray}
The quantities $W_{ij}$ denote the rates of the transitions $i \rightarrow j$ allowed by the dynamics.
The master equation can be written in matrix form as
\begin{eqnarray}
\frac{d \boldsymbol p (t)}{dt} = \boldsymbol p(t) \hat{L}
\end{eqnarray}
where we introduced the operator $\hat{L}$ with elements $\hat{L}_{ij} = W_{ij}$ for $i \neq j$ and $\hat{L}_{ii} = - \sum_j W_{ij}$ otherwise.
Under general assumptions \cite{S76} the system evolves towards a unique stationary state $\boldsymbol p_{\rm st}$ satisfying $d \boldsymbol p_{\rm st}/dt = 0$. 

The concept of {\it uniformization} of a Markov process has been introduced to help the sudy of such continuous-time random processes \cite{J53}. It transforms the continuous-time process into another system evolving in {\it discrete time} while preserving many key properties of the dynamics. The construction proceeds as follows.

Introducing the inverse time step 
\begin{eqnarray}
\beta \geq \max_i |\hat{L}_{ii}|
\label{beta}
\end{eqnarray}
we define the transition matrix $\hat{U}(\beta)$ by $\hat{U}_{ij} = W_{ij}/\beta$ for $i \neq j$, and $\hat{U}_{ii} = 1 - \sum_j W_{ij}/\beta$ otherwise. In matrix form it reads
\begin{eqnarray}
\hat{U} (\beta) \equiv \hat{I} + \frac{\hat{L}}{\beta}
\end{eqnarray}
where $\hat{I}$ is the identity matrix. It is readily verified that $\hat{U}(\beta)$ is a proper transition matrix, i.e. $\sum_j \hat{U}_{ij} = 1$ and $\hat{U}_{ij}\geq 0$ for all $i,j$ and all $\beta  \geq \max_i |L_{ii}|$. The uniformized Markov chain thus evolves over the same state space $\{ i \}$ but in discrete time steps of size $\Delta t=1/\beta$, with the correspondance $t= n \times \Delta t = n/\beta$. For simulation purposes the time step $\Delta t$ should be chosen as large as possible, so that the optimal value of $\beta$ satisfies the equality in Eq. (\ref{beta}).

The probability distribution $\boldsymbol \pi_\beta (n)$ over the uniformized system evolves according to the discrete-time evolution equation
\begin{eqnarray}
\boldsymbol \pi_\beta (n) = \boldsymbol \pi_\beta (n-1) \hat{U}(\beta) \, .
\label{pievol}
\end{eqnarray}
Remarkably, the stationary state $\boldsymbol \pi_{\rm st}$ of the uniformized system exactly corresponds to the stationary state $\boldsymbol p_{\rm st}$ of the original system: $\boldsymbol p_{\rm st} \hat{U}(\beta) =  \boldsymbol p_{\rm st} (\hat{I}+\hat{L}/\beta) = \boldsymbol p_{\rm st}$ so that $\boldsymbol \pi_{\rm st} = \boldsymbol p_{\rm st}$ for all values of $\beta$. 

The uniformization procedure thus provides a discrete-time formulation of the original dynamics that preserves the steady state distribution. It is not, however, an exact mapping of the dynamics as several other properties may depend on the parameter $\beta$ (e.g., the topological entropy). In the next section we show how to relate the large deviation functions obtained from the original continuous-time process to those obtained from the uniformized system.\\

\section{Large deviation functions from uniformized dynamics}

Large deviation functions play an increasingly important role in many different fields \cite{T09}. They describe the occurrence of rare events, that is the large fluctuations away from the mean behavior. 
Recent developments have highlighted symmetry properties in the large fluctuations of far-from-equilibrium thermodynamic quantities \cite{SPWS08}.
Here we study two such variables: the dissipation rate and the thermodynamic currents.
We show that, in both cases, the generating functions obtained in continuous time can be exactly recovered from those arising in the discrete-time domain.

We first consider the dissipation rate $S(t)$. $S(t)$ is a fluctuating quantity measuring the dissipation occurring along a specific trajectory of the system. In the present context 
\begin{eqnarray}
S(t) \equiv  \ln \parent{ \prod_{{\rm traj}}  W_{ij}/W_{ji} } \, .
\end{eqnarray}
Its large deviation function $I$ is defined as \cite{V84}
\begin{eqnarray}
{\rm Prob}[ S(t)/t = \xi] \sim e^{-t I(\xi)} \qquad (t \rightarrow \infty) \, .
\label{ldf}
\end{eqnarray}
Instead of studying the large deviation function (\ref{ldf}) directly, we develop our analysis at the level of the generating function, which is defined via the Legendre transform $G(\lambda) = \max_\xi [I(\xi)-\xi\lambda]$.
Alternatively, the generating function associated with equation (\ref{ldf}) can be expressed as the limit
\begin{eqnarray}
G(\lambda)  = \lim_{t\rightarrow \infty} - \frac{1}{t} \ln \left \langle e^{-\lambda S(t)} \right \rangle \, .
\label{Q}
\end{eqnarray}
The generating function allows us to obtain all cumulants of the dissipation by taking successive derivative with respect to $\lambda$: $\cum{S^n} = (-1)^{n-1} d^n G/d\lambda^n (0)$. The average in Eq. (\ref{Q}) is calculated as $\left \langle e^{-\lambda S(t)} \right \rangle = ||  \boldsymbol g_\lambda(t) ||_1$, where $|| x ||_1 = \sum_i |x_i|$ is the $L_1$-norm. The vector $\boldsymbol g_\lambda(t)$ satisfies the initial condition $\boldsymbol g_\lambda(0) = \boldsymbol p(0)$ for all $\lambda$ and 
evolves according to \cite{LS99} 
\begin{eqnarray}
\frac{d\boldsymbol g_\lambda(t)}{dt} = \boldsymbol g_\lambda(t) \hat{L}_\lambda 
\label{dFdt}
\end{eqnarray}
with the operator $\hat{L}_\lambda$ given by
\begin{equation} 
\hat{L}_\lambda = \begin{cases}
                 W^{1-\lambda}_{ij} W^{\lambda}_{ji} &  \text{if $i \neq j$} \\
                 - \sum_j W_{ij}  & \text{otherwise}.
                 \end{cases} \nonumber
\end{equation}
Accordingly, the generating function (\ref{Q}) is given by minus the largest eigenvalue of the operator $\hat{L}_\lambda$. 
Note that for $\lambda=0$ we recover the evolution operator $\hat{L}=\hat{L}_{\lambda = 0}$ for the probability distribution $\boldsymbol p(t)$.

We now consider the uniformized process (\ref{pievol}). The analogue of the generating function (\ref{Q}) is defined as
\begin{eqnarray}
\bar{G}(\lambda) = \lim_{n\rightarrow \infty} - \frac{1}{n} \ln  \left \langle e^{-\lambda \bar{S}(n)}  \right \rangle
\label{Q'}
\end{eqnarray}
with $\bar{S} = \ln \parent{ \prod_{{\rm traj}} U_{ij}/U_{ji} }= \ln \parent{ \prod_{{\rm traj}} W_{ij}/W_{ji} } = S$, which now evolves in discrete time. Similarly, $\bar{G}$ can be obtained as minus the largest eigenvalue of the operator
\begin{equation}
\hat{U}_\lambda = \begin{cases}
                 (1/\beta) W^{1-\lambda}_{ij} W^{\lambda}_{ji} & \mbox{if $i \neq j$} \\
                 1- (1/\beta) \sum_j W_{ij} & \mbox{otherwise}
                 \end{cases} \nonumber
\end{equation}
or
\begin{eqnarray}
\hat{U}_\lambda = \hat{I} + \frac{\hat{L}_\lambda}{\beta} \, .
\label{UtoL}
\end{eqnarray}
In turn we recover the generator of the time evolution when $\lambda = 0$:  $\hat{U}=\hat{U}_{\lambda = 0}$. 

To establish the connection between the generating functions (\ref{Q}) and (\ref{Q'}), we derive the explicit relation between the eigenvalues $\mu_n$ and $\bar{\mu}_n$ of the two processes. 
Using the relation (\ref{UtoL}) between the original and the uniformized process, we see that the eigenvalue equation reads 
\begin{eqnarray}
\det [ \hat{L}_\lambda - \mu_n \hat{I} ] =  \det [\beta \hat{U}_\lambda -\beta \hat{I}- \mu_n \hat{I}] =  0 
\end{eqnarray}
or 
\begin{eqnarray}
\det[  \hat{U}_\lambda - \hat{I} (\mu_n/\beta+1)  ] = 0 \, .
\end{eqnarray}
This last expression reveals that if $\mu_n$ is an eigenvalue of the original operator $\hat{L}_\lambda$,  then $\bar{\mu}_n= \mu_n/\beta +1$ is an eigenvalue of the discrete-time evolution operator $\hat{U}_\lambda$. Therefore all eigenvalues are simply scaled by a factor $\beta$ and shifted by the unity. In addition, all eigenvectors can be verified to be strictly identical between the two processes \cite{F00}. We thus arrive at our main result: The generating function (\ref{Q'}) of the uniformized system is related to the original generating function (\ref{Q}) by the linear transformation
\begin{eqnarray}
\bar{G}(\lambda) = G(\lambda)/\beta+1.
\label{Qscale}
\end{eqnarray}
The corresponding large deviation functions are thus related through the scaling $\bar{I}(\xi) = I(\xi\beta)/\beta$. 
This result demonstrates that the large fluctuations can be exactly obtained from the discrete-time dynamics of the uniformized process. This simplifies many theoretical and numerical formulations, as will be discussed in the next sections. 

We next consider the thermodynamic currents, which measure the transport of matter and energy inside the system and their exchanges with the environment. They are expressed as 
\begin{eqnarray}
J (t) = \sum_{{\rm traj}} \epsilon (t) \delta(t-t_{{\rm jump}})\, ,
\label{J.def}
\end{eqnarray} 
where $\epsilon = \pm 1$ if a transition contributes to the current in the positive or negative direction, respectively, and zero otherwise \cite{AG} (see also Section \ref{sec_example}). 
A similar derivation can be obtained for the generating function $Q$ of the currents, leading to $\bar{Q} = Q/\beta +1$. This conclusion stems from the observation that the physical quantities of interest (entropy production, thermodynamic currents) do not affect the diagonal terms of the corresponding large deviation operators (no entropy and no currents are generated when no jumps occur). 

This strong correspondence is unanticipated. 
Indeed, the generating function (\ref{Q}) is equivalently expressed as a path integral over all possible trajectories:
\begin{eqnarray}
G =\lim_{t\rightarrow \infty}  - \frac{1}{t} \ln \int_{{\rm traj}} P({\rm traj}) \exp [-\lambda S({\rm traj})]. 
\end{eqnarray}
The probability of a trajectory reads 
\begin{eqnarray}
P({\rm traj}) = p_0 \prod_i \exp[(t_{i+1}-t_i)\hat{L}_{ii}] W_{ii+1}
\end{eqnarray}
and depends on the exact transition times $(t_1, t_2, \cdots, t_n, \cdots)$. Thus, in principle, the large deviations should reflect the fine temporal structure (the time intervals between jumps, weighted by the corresponding factors $\hat{L}_{ii}$) of the continuous-time process. Now, in discrete-time, the probability of the same path reads $\pi_0 \prod_i \hat{U}_{ii+1}(\beta)$, irrespective of its temporal structure, while the dissipation is identical in both formulations ($\bar{S}=S$). Yet, the result (\ref{Qscale}) shows that, in the discrete-time domain, the generating function is simply scaled by the discretization parameter $\beta$. In this sense, all information on the large deviations are contained in the discrete-time dynamics (\ref{pievol}). 
More generally, all eigenvalues being closely related to those of the original process, we expect the continuous- and discrete-time dynamics to present strong connections. We explore this issue in the next section.\\

\section{Finite-time dynamics and simulation algorithms}
\label{finitetimedynamics}

In the previous section we demonstrated that the study of the large deviation functions can be performed using the discrete-time dynamics (\ref{pievol}). These functions are defined in the infinite-time limit; here we further develop the link between the two descriptions and analyze the finite-time regime. 

The detailed connection with the continuous-time evolution is accomplished through the following construction.
The solution of the system (\ref{dFdt}) can be written as
\begin{eqnarray}
\boldsymbol g_\lambda (t) &=& \boldsymbol p (0) e^{\hat{L}_\lambda t}  \nonumber \\
&=& \boldsymbol p (0)  e^{- \beta t \hat{I} + \beta t \hat{U}_\lambda(\beta)} \nonumber \\
&=& e^{- \beta t} \sum_{k=0}^\infty \frac{(\beta t)^k}{k!}  \boldsymbol p (0) \hat{U}_\lambda^k(\beta) \, ,
\label{finitet}
\end{eqnarray}
where we used that $\boldsymbol g_\lambda(0) = \boldsymbol p(0)$ in the first line, the relation (\ref{UtoL}) in the second line, and the commutativity of the identity operator in the third line.
The last expression provides a robust way of numerically evaluating the finite-time generating functions \cite{F01}. Indeed, as opposed to the operator $\hat{L}_\lambda$, the operator $\hat{U}_\lambda$ and its powers $\hat{U}_\lambda^k$ have non-negative elements only.
This in turn implies that the last expression has no additions of numbers with opposite signs. This is an advantageous feature because additions of numbers with opposite sign increase round-off errors considerably \cite{G77}. 

A revealing interpretation of the previous formula can be gained from the following consideration. Recall that the time evolution of the probability distribution $\boldsymbol p(t) = \boldsymbol g_{\lambda  =0} (t)$ is recovered in the special case $\lambda  = 0$. Introducing the number of transitions $N(t)$ during the time interval $[0,t]$, equation (\ref{finitet}) with $\lambda  = 0$ can be interpreted as 
\begin{eqnarray}
{\rm Prob} [j, t] = \sum_{k, i} {\rm Prob} [j,t | i,0 \; {\rm and} \; N(t)=k] \times {\rm Prob}[N(t)=k]
\label{proba}
\end{eqnarray}
where
\begin{eqnarray}
{\rm Prob}[j,t | i,0 \; {\rm and} \; N(t)=k] = (\hat{U}^k)_{ij}(\beta) 
\label{transfer}
\end{eqnarray}
is the probability to be in state $j$ at time $t$ given the initial state $i$ at time $t=0$ and a number of transitions $k$ occurring during the time interval $t$. The probability to observe a number of transitions $N(t) = k$ during the time interval $t$ reads
\begin{eqnarray}
{\rm Prob}[N(t) = k] = e^{- \beta t} \frac{(\beta t)^k}{k!} \, ,
\label{poisson}
\end{eqnarray}
i.e. it satisfies a Poisson process of mean $\beta t$. This formulation is remarkable for it implies that we can express the original stochastic process in terms of a single homogeneous Poisson rate. Alternatively, this Poisson distribution can be interpreted as arising from the sum of independent exponential distributions. In this case the system jumps to another state $j$ with probability $\hat{U}_{ij}(\beta)$ after a random waiting time exponentially distributed with mean $1/\beta$, regardless of the current state $i$ (hence the name uniformization).

Building on this interpretation, we can devise alternative strategies to compute quantities of interest such as the transition probabilities $\boldsymbol p(t)$ or the generating function $\boldsymbol g_\lambda (t)$. Indeed, we can sample the space of trajectories using the following algorithm:\\

{\bf Simulation algorithm for autonomous processes}\\

(1) Generate a random number $N$ sampled from a Poisson distribution of mean $\beta t$.

(2) Generate a random trajectory of length $N$ according to the discrete-time process $\hat{U}(\beta)$.\\

Equations (\ref{proba})-(\ref{poisson}) ensure that this construction generates the correct probability distribution. Note that the present scheme completely bypasses the random exponential waiting times needed in the traditional formulation \cite{G76} while remaining exact. Equation (\ref{finitet}) also guarantees that, for $\lambda \neq 0$, using these randomly generated trajectories leads to the exact generating function. 
In addition to its simplicity, the present algorithm uses an mean number of random numbers equal to $1+\beta t$ to generate a trajectory of length $t$. In contrast, Gillespie's algorithm \cite{G76} requires $2 t \mean{L}$ random numbers in average, where $\mean{L} = (1/t) \sum_i  |\hat{L}_{ii}| \int_0^t p_i (\tau)d\tau$ is the mean waiting time between jumps. Accordingly, when the inverse time step $\beta$ can be chosen such that $\beta < 2 \mean{L}$ the present formulation is expected to outperform Gillespie's algorithm.

Importantly, this algorithm can be generalized to encompass {\it time-dependent} Markov processes. Consider a time-dependent system described by an evolution operator $\hat{L}(t)$ and choose $\beta \geq \max_{i,t} |\hat{L}_{ii}(t)|$. Then the space of trajectories can be sampled by iterating the following steps:\\ 

{\bf Simulation algorithm for non-autonomous processes}\\

(1) Generate a waiting time $\tau$ exponentially distributed with a mean $1/\beta$, independently of the current state $i$. 

(2) Update the current time: $t \leftarrow t+\tau$.

(3) Jump to a state $j$ randomly selected with probability $\hat{U}_{ij}(\beta,t)$.\\ 

The total number of jumps is also given by the Poisson process (\ref{poisson}), but here it is necessary to keep track of their exact timings due to the time dependence of the system. Note that the optimal value of $\beta$ for simulation purposes is given by $\beta = \max_{i,t} |\hat{L}_{ii}(t)|$ as it minimizes the number of steps needed to generate a trajectory. It can thus be directly estimated from the knowledge of the transition rates.

Remarkably this representation avoids the need to consider inhomogeneous, {\it time-dependent} waiting times. This feature is especially important as generating random numbers according to a distribution of the form $|\hat{L}_{ii}(t)| \exp[\int_{t_0}^t \hat{L}_{ii}(\tau)d\tau]$ is difficult and approximate in most situations of interest (see next section). In contrast, step $1$ only requires exponential random numbers. 
As shown in the Appendix, the same construction applies to the generating functions as well.
We illustrate these exact simulation methods in the next section.

\section{Example: fluxes in stochastic pumps}
\label{sec_example}

In this section we illustrate our results on a model of stochastic pumping. Such models play an increasingly important role, e. g., in the study of molecular motors. They remain, however, very difficult to simulate stochastically due to their time dependence.

\begin{figure}[t]
\centerline{\includegraphics[width=8cm]{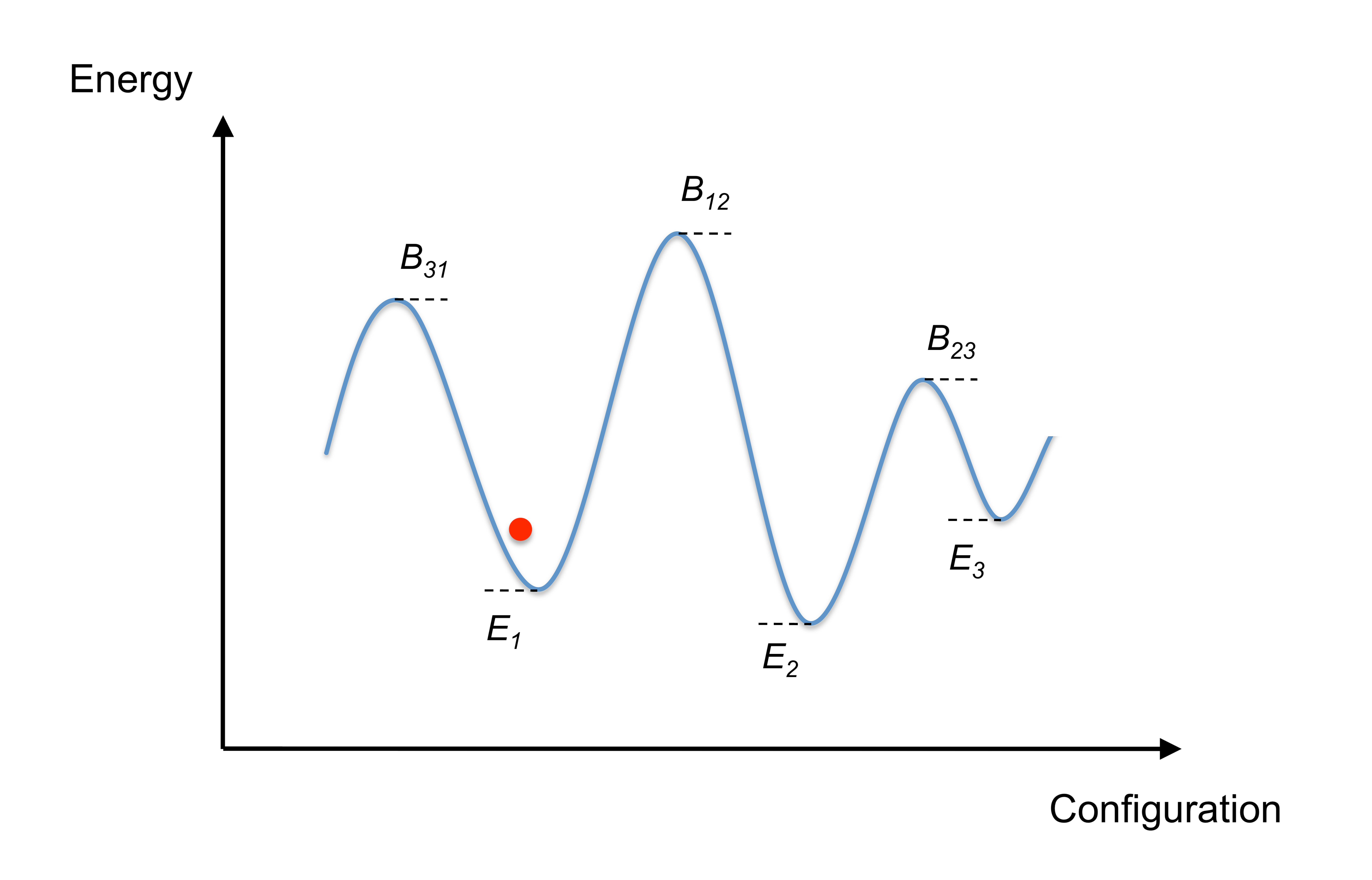}}
\caption{(Color online) A model of stochastic pump. The particle makes thermal transitions among three states with energies $E_i$, over barriers with energies $B_{ij}$. The temperature is varied in time to induce currents.}
\label{fig1}
\end{figure}

We consider a model system motivated by an experiment by Leigh et. al. \cite{Leigh03} and analyzed by Astumian \cite{A07} and Rahav et al. \cite{RHJ08}.
The system consists in thermally activated transitions among three states, depicted by the wells and energy levels in Fig. \ref{fig1}, with rates $W_{ij} = k e^{-\beta[B_{ij}-E_i]}$. We will take $k,\beta=1$ to set the units of time and energy. The system satisfies the Kolomogorov condition $W_{12}W_{23}W_{31}=W_{13}W_{32}W_{21}$ and is thus assumed to be at equilibrium initially. Here we induce non-zero currents by periodically varying the temperature of the system:
\begin{eqnarray}
\beta(t) = 1 + A \sin \parent{2\pi \frac{t}{T}} \, .
\label{betaoft}
\end{eqnarray}
We consider the pumped flux
\begin{eqnarray}
\Phi(t) = \int_0^t J(\tau) d\tau \,
\label{Phi}
\end{eqnarray}
induced by this temperature variation. 
Its fluctuations can be described by the moment generating function 
\begin{eqnarray}
F_\lambda (t)  = \left \langle e^{-\lambda \Phi(t) } \right \rangle \, .
\label{F}
\end{eqnarray}
All moments of the pumped flux distribution can be obtained by calculating derivatives with respect to $\lambda$: 
\begin{eqnarray}
\mean{\Phi^n(t)} = (-1)^{n} \frac{d^n F(t)}{d\lambda^n} \Big\vert_{\lambda = 0} \, .
\end{eqnarray}
The generating function (\ref{F}) can be expressed as $F_\lambda= || \boldsymbol f_\lambda ||_1$ in terms of the vector $\boldsymbol f_\lambda$ satisfying
\begin{eqnarray}
\frac{d\boldsymbol f_\lambda(t)}{dt} = \boldsymbol f_\lambda(t) \hat{H}_\lambda(t) 
\label{dfdt}
\end{eqnarray}
where 
\begin{equation} 
\hat{H}_\lambda (t) = \begin{cases}
                 W_{ij} (t) e^{-\lambda \epsilon_{ij}} &  \text{if $i \neq j$} \\
                 - \sum_j W_{ij} (t) & \text{otherwise}.
                 \end{cases} \nonumber
\end{equation}
The quantity $\epsilon_{ij} = -\epsilon_{ji}$ takes the value $\pm 1$ if the transition $i \rightarrow j$ generates a positive (negative) current, and zero otherwise \cite{AG}. We recover the time evolution operator for $\lambda=0$: $\hat{L} = \hat{H}_{\lambda=0}$.

\begin{figure}[t]
\centerline{\includegraphics[width=9cm]{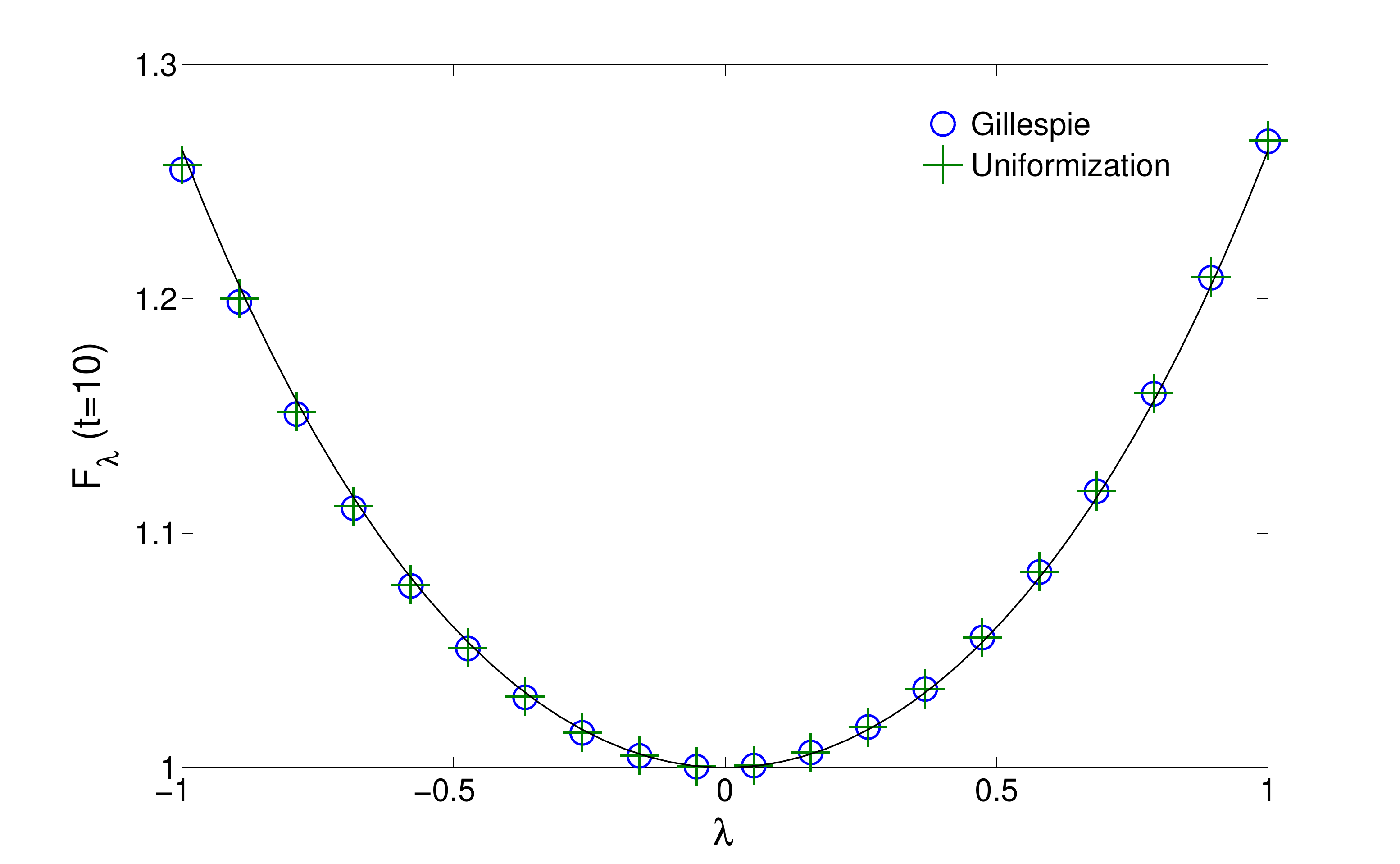}} 
\caption{(Color online) Fluctuations of the integrated flux $\Phi_{12}(t)= \int_0^t J_{12}(\tau) d\tau$ at equilibrium (i.e., without driving), as measured by its moment generating function (\ref{F}). The current is measured in terms of $\epsilon_{12} = - \epsilon_{21} = 1$ and zero otherwise. The well depths take the values $(E_1, E_2, E_3) = (-1.5, -1.75, -2.25)$ and the barriers $(B_{12}, B_{23}, B_{13}) =(-0.3, 0.5, 0)$. The solid line denotes the numerical solution of the system (\ref{dfdt}). 
The circles and pluses correspond to the simulation of 50000 trajectories using Gillespie's and the uniformization algorithm, respectively.}
\label{fig2}
\end{figure}

We first perform stochastic simulations in the equilibrium state without time-dependent driving. The solid line in Fig. \ref{fig2} shows the moment generating function of the integrated current $\Phi_{12} (t=10) = \int_0^{t=10} J_{12}(\tau) d\tau$, obtained by numerical integration of Eq. (\ref{dfdt}). The circles and pluses denote the results of 50000 random trajectories of length $t=10$, sampled according to Gillespie's and the uniformized algorithm, respectively. Being exact, both approaches present an excellent agreement with the solution of the system (\ref{dfdt}).
The average number of random numbers needed is, however, different. As discussed in the previous section, we expect the uniformized dynamics to outperform Gillespie's algorithm when the waiting times have the same order of magnitude. For this set of parameters, the ratio between the largest and the lowest waiting time is around $3$. Yet, Gillespie's algorithm requires $2\mean{L} \approx 0.5996$ random numbers per unit time and per trajectory, which is larger than $\beta \approx 0.5335$ for the uniformized dynamics.

\begin{figure}[t]
\centerline{\includegraphics[width=9cm]{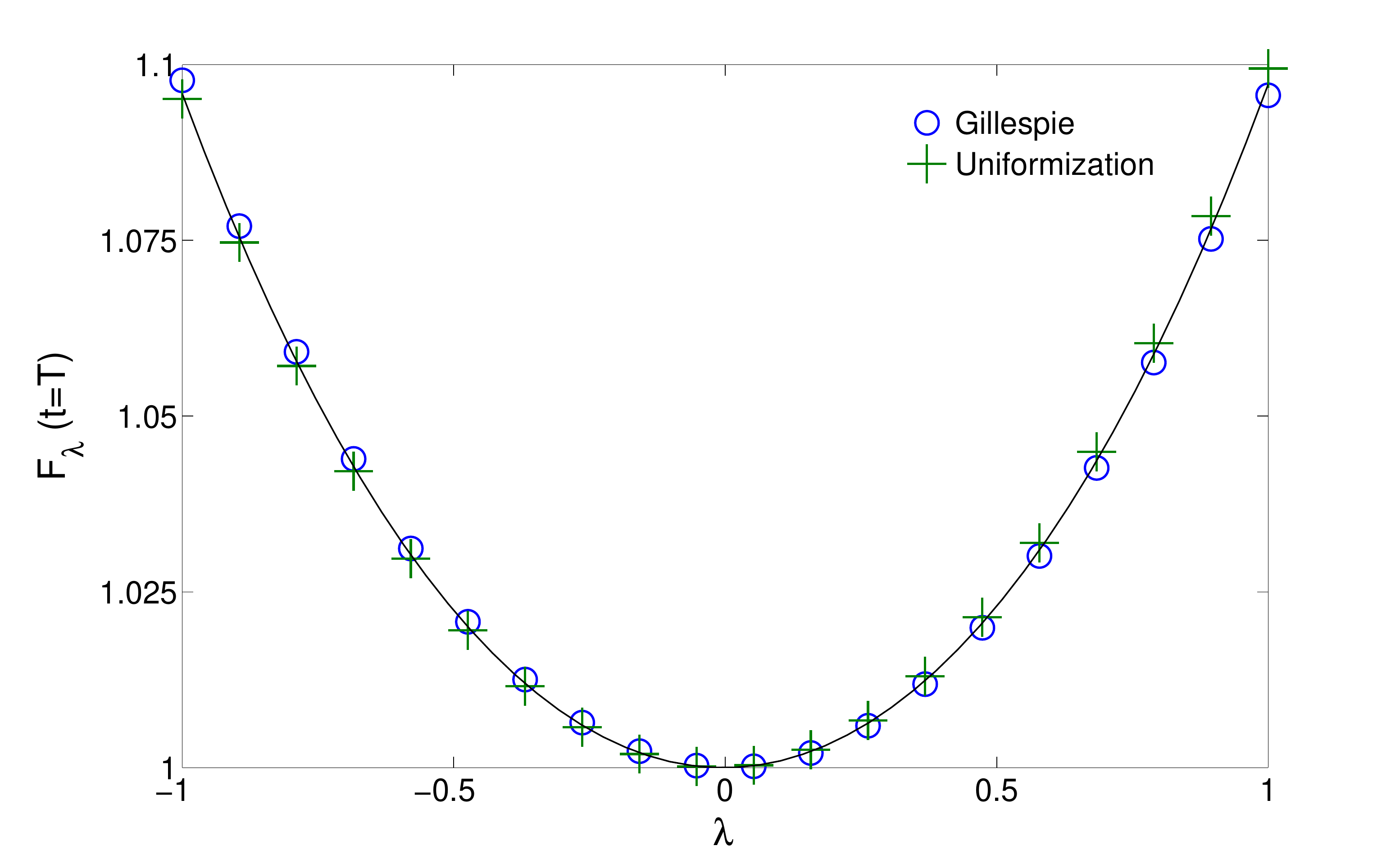}}
\caption{(Color online) Fluctuations of the integrated flux $\Phi_{12} (T)= \int_0^T J_{12}(\tau) d\tau$ over one cycle of the temperature driving (\ref{betaoft}), as measured by its moment generating function (\ref{F}). The driving period $T=2$ and its amplitude $A=0.1$. All other parameters are identical to those in Fig. \ref{fig2}. The solid line denotes the numerical solution of the system (\ref{dfdt}). 
The circles and pluses correspond to the simulation of 50000 trajectories using Gillespie's and the uniformization algorithm, respectively.}
\label{fig3}
\end{figure}

We now turn to the situation in presence of the time-dependent driving (\ref{betaoft}), for which we observe a drastic difference between the two approaches.
We implemented the simple time-dependent uniformized algorithm described in Section \ref{finitetimedynamics}, while the time-dependent Gillespie algorithm was implemented as follows. When the system is in state $i$ at time $t$, a random transition time $\tau > t$ is generated according to the distribution
\begin{eqnarray}
R_{i}(t; \tau) &=& |\hat{L}_{ii}(\tau)| P_i(t;\tau)  \nonumber \\
     &=&  |\hat{L}_{ii}(\tau)| \exp \parent{\int_{t}^\tau \hat{L}_{ii}(s)ds}\, .           
     \label{Roft}
\end{eqnarray}
To sample this {\it time-dependent} distribution, we generate a uniform random number $r$ between $[0,1]$ and solve the equation $P_i(t;\tau)-r=0$. Note that finding the zero of this equation involves a finite number of evaluation of the functions $P$, which in turn implies the evaluation of that many integrals. 
Finally, we update the current time ($t \leftarrow \tau$) and select a new state $j \neq i$ with probability $\hat{L}_{ij}(t)/|\hat{L}_{ii}(t)|$ .

\begin{table}[t]
\centerline{
\begin{tabular}{|c|c|c|}
\hline
& Gillespie & Uniformization \\\hline
 Random numbers &  0.3 & 0.6 \\
 Integrals &  28.09 & 0 \\
 Simulation time &  321.18 & 1 \\
\hline
\end{tabular}
}
\caption{Computational cost (per trajectory and per unit time) for the simulation of Fig. \ref{fig3}. 
In the time-dependent case the number of random numbers is given by $(2/t) \sum_i \int_0^t  p_i (\tau) |L_{ii}(\tau)| d\tau$ for Gillespie's algorithm and by $2\beta$ for the uniformized algorithm.}\label{table1}
\end{table}

We show in Figure \ref{fig3} the generating function $F_\lambda(T)$ after one cycle. Here also, both algorithms display a similar degree of accuracy. Their computational cost is, however, very different, as revealed in Table \ref{table1}. Although the uniformized algorithm requires more transitions in average, Gillespie's algorithm requires the generation of random numbers following distributions of the form (\ref{Roft}), which is a computationally intensive task. For this reason, the uniformized algorithm was running $>300$ times faster than Gillespie's algorithm \cite{matlab}. Changing the parameters always lead to comparable improvements.

A similar improvement is expected for many non-autonomous systems. In such systems, the computational bottleneck is the generation of random numbers following time-dependent distributions of the form (\ref{Roft}). This requires finding the zero of an equation, whose evaluation requires the calculation of integrals, for {\it each} transition. In contrast the uniformized algortithm only requires exponential random numbers, providing a more straightforward and efficient implementation.

\section{Conclusions}

We have described a mapping from general continous-time Markov processes to discrete-time Markov chains that presents several key features. 
First, all eigenvectors of the original dynamics, including the steady state, are strictly preserved.  
Second, all eigenvalues are related by a linear transformation.
Third, it offers an interpretation of the time evolution in terms of a single homogeneous Poisson rate (uniformization). 

We have demonstrated that this uniformization procedure also preserves the generating functions of the original process. 
In particular, we have analyzed the generating functions of the dissipation rate and of the thermodynamic currents. More generally, this conclusion will hold true for all physical quantities that only vary during the transitions between states of the system.
Although the generating functions evolve according to generalized operators that do not present a transition matrix structure, the time-discrete dynamics preserves the associated generalized eigenvectors while the eigenvalues are related via a linear transformation. Thus, for the purpose of studying generating functions, it is sufficient to focus on the uniformized dynamics exclusively.

This framework provides important simplifications for the theoretical and numerical analysis of large deviation functions. 
In particular, it allows the implementation of efficient numerical techniques to study the dynamics and fluctuations of Markov processes. 
We have illustrated the derived simulation algorithms on a model of stochastic pumping. The fluxes pumped by a time-dependent driving have a considerable importance in many applications but their stochastic simulation has remained a challenge. As we have shown, the present approach offers important advantages both at the level of simplicity and efficiency, especially, for non-autonomous systems. Remarkably, we have observed a two-order of magnitude improvement in the simulation of a stochastic pump. This approach thus provides a powerful tool to study ratchets and pumps \cite{S09} and, more generally, all time-dependent systems such as temperature-programmed desorption experiments \cite{J95} or driven quantum dots \cite{FHS06}.  

Theoretical insights can also be gained from this formulation. For instance, it reveals that the temporal aspect of continous-time trajectories - even for inhomogeneous processes - does not contain information on the large deviation functions. 
To explore the scope of this conclusion, the next natural step is to consider semi-Markovian processes for which the waiting times between jumps exhibit arbitrary distributions. In this case the generating function is given by the solution of an equation involving the Laplace transforms of the waiting time distributions \cite{AG08}. However, when the waiting times are not exponentially distributed (i.e., the non-Markovian case), this equation cannot be written as an eigenvalue problem any longer. As a result, even tough some uniformization procedure can be derived in this case as well \cite{S80b}, the generating functions cannot be obtained from any discrete-time dynamics. In this situation the generating functions are shaped by the precise form of the waiting time distributions, even for homogeneous processes. The present approach thus provides a systematic way to disentangle the contributions of the non-Markovianity to the large deviation functions.\\

\section{Appendix}
In this appendix we extend the uniformization procedure for non-autonomous processes \cite{vD92} to the case of the generating function operator. We consider a time-dependent Markov process characterized by an evolution operator $\hat{L}(t)$. The vector $\boldsymbol g_\lambda(t)$ evolves according to
\begin{eqnarray}
\frac{d\boldsymbol g_\lambda(t)}{dt} = \boldsymbol g_\lambda(t) \hat{L}_\lambda(t) 
\end{eqnarray}
with 
$\hat{L}_\lambda (t)=  W^{1-\lambda}_{ij} (t) W^{\lambda}_{ji} (t)$ if $i \neq j$ and $- \sum_j W_{ij}(t)$  otherwise.
Due to the non-commutativity of the operator $\hat{L}(t)$ at different times, the solution to this evolution equation is expressed as the Peano-Baker series \cite{I56}
\begin{eqnarray}
\lefteqn{\boldsymbol g_\lambda(t) = \boldsymbol p(0) \sum_{k=0}^{\infty} \int_0^t ds_1 \int_{s_1}^t ds_2 \cdots} \nonumber\\ 
& &\times \int_{s_{k-1}}^t ds_k \, \hat{L}_\lambda(s_1) \hat{L}_\lambda(s_2) \cdots \hat{L}_\lambda(s_k) \, . \
\label{pbaker}
\end{eqnarray}
Introducing $\beta \geq \max_{i,t} |\hat{L}_{ii}(t)|$ the operator $\hat{U}(t) = \hat{I} +  \hat{L}(t)/\beta$ defines a transition matrix at all times. Its continuation for $\lambda \neq 0$ reads
\begin{eqnarray}
\hat{U}_\lambda(t) = \hat{I} + \frac{\hat{L}_\lambda(t)}{\beta}
\label{uoft}
\end{eqnarray}
and describes the evolution of the moment generating function. Substituting formula (\ref{uoft}) into the series (\ref{pbaker}) and using that
\begin{eqnarray}
\int_0^t ds_1 \int_{s_1}^t ds_2 \cdots  \int_{s_{k-1}}^t ds_k = \frac{t^k}{k!} \, ,
\end{eqnarray}
we obtain, after some manipulations, 
\begin{eqnarray}
\lefteqn{ \boldsymbol g_\lambda(t) =\boldsymbol  p(0) e^{- \beta t} \sum_{k=0}^{\infty} \frac{(\beta t)^k}{k!} \int_0^t ds_1 \int_{s_1}^t ds_2 \cdots}  \nonumber \\ 
& & \times  \int_{s_{k-1}}^t ds_k  \left( \frac{k!}{t^k}\right) \hat{U}_\lambda(s_1) \hat{U}_\lambda(s_2) \cdots \hat{U}_\lambda(s_k) \, . \quad
\label{glpeano}
\end{eqnarray}
This expression has the following interpretation. The terms $e^{-\beta t}(\beta t)^k/k!$ are Poisson probabilities. The integration accounts for all possible sets of time points at which events in the Poisson process can take place. The term $k!/t^k$ corresponds to the density introduced by the Poisson process, for which the probability density of $k$ events is uniformly distributed in the interval $[0,t]$. Expressing the Poisson probabilities as the sum of independent exponential random times of mean $1/\beta$ we deduce the simulation algorithm presented in main text. Moreover, expression (\ref{glpeano}) implies that the generating function will be adequately sampled.

\acknowledgments
This work is supported by the F.~N.~R.~S. Belgium. We thank two anonymous referees for their constructive comments.


\begin{thebibliography}{0}
\bibitem{F00} If $\boldsymbol e$ is an eigenvector of the operator $\hat{L}_\lambda$, i.e. $\hat{L}_\lambda \boldsymbol e = \mu \boldsymbol e $, then it is also an eigenvector of the uniformized operator $\hat{U}_\lambda$: $\hat{U}_\lambda \boldsymbol e = (\hat{I} + \hat{L}_\lambda/\beta)\boldsymbol e = (1+\mu/\beta) \boldsymbol e$. The converse is also true.
\bibitem{F01} A stronger statement can be formulated. Denoting the largest eigenvalue of $\hat{U}_\lambda (\beta)$ by $q_\lambda$, its associated eigenvector by $\boldsymbol e_\lambda$, and $c_\lambda = \boldsymbol e_\lambda \cdot \boldsymbol p(0)$, the approximation
\begin{eqnarray}
\boldsymbol g_\lambda (t) \approx  e^{q_\lambda t} c_\lambda \, \boldsymbol e_\lambda  + e^{- \beta t} \sum_{k=0}^N \frac{(\beta t)^k}{k!}  \Big[\boldsymbol p(0) \hat{U}_\lambda^k(\beta)- \, \boldsymbol e_\lambda q^k_\lambda c_\lambda \Big] 
\end{eqnarray}
can be shown to be uniform in $t$, that is the truncation level $N$ can be chosen such that the approximation has a specified accuracy for all $t\geq 0$ {\it simultaneously}.
\end{thebibliography}
\end{document}